\documentclass[12pt,twoside,a4paper]{article}
\pdfoutput=1

\usepackage{amsmath, amsthm, amsfonts, amssymb}
\usepackage{graphicx}
\usepackage{multirow}
\usepackage{float}
\usepackage{color}
\usepackage{cite}
\usepackage[width=\textwidth,font=small,labelfont=bf,
labelsep=endash]{caption}
\usepackage{titlesec}

\titleformat{\section}
  {\normalfont\fontsize{13}{16}\bfseries}{\thesection}{1em}{}
\titleformat{\subsection}
  {\normalfont\fontsize{11}{14}\bfseries}{\thesubsection}{1em}{}

\normalsize
\newcommand\csch            {\mathrm{csch}}

\newcommand\doi[2]        {\href{http://dx.doi.org/#1}{#2}}

\makeatletter
\newcommand{\vast}{\bBigg@{4}}
\newcommand{\Vast}{\bBigg@{5}}
\makeatother

\begin{document}

\title{\textbf{Area Law Behaviour of Mutual Information at Finite Temperature}}
\author{Dimitrios Katsinis$^{1,2}$ and Georgios Pastras$^2$}
\date{\small $^1$Department of Physics, National and Kapodistrian University of Athens,\\University Campus, Zografou, Athens 15784, Greece\\
$^2$NCSR ``Demokritos'', Institute of Nuclear and Particle Physics,\\Aghia Paraskevi 15310, Attiki, Greece\linebreak \vspace{8pt}
\texttt{dkatsinis@phys.uoa.gr, pastras@inp.demokritos.gr}}

\vskip .5cm

\maketitle

\begin{abstract}
Entanglement entropy in free scalar field theory at its ground state is dominated by an area law term. However, when mixed states are considered this property ceases to exist. We show that in such cases the mutual information obeys an ``area law''. The proportionality constant connecting the area to the mutual information has an interesting dependence on the temperature. At infinite temperature it tends to a finite value which coincides with the classical calculation.
\end{abstract}

\newpage

%

\setcounter{equation}{0}
\section{Introduction}
\label{sec:intro}

Entanglement entropy in free scalar field theory at its ground state is dominated by an area law term \cite{Srednicki:1993im}. This is an intriguing similarity with black hole physics. The same similarity has been independently discussed by Bombelli et.al. \cite{Bombelli:1986rw}, who calculated the entanglement entropy in scalar field theory at a black hole background. From a quantum mechanical point of view, the dependence of the entanglement entropy on the geometric characteristics of the entangling surface, and not on those of the subsystem itself, is a natural consequence of the symmetric property of the entanglement entropy $S_A = S_{A^C}$, where $A^C$ is the subsystem complementary to the subsystem $A$. As an indicative example, the symmetry property excludes the existence of a volume term; this term should be proportional to the volume of the subsystem under consideration and at the same time proportional to the volume of its complement.

However the symmetry property of entanglement entropy holds only for pure states of the overall system. When mixed, e.g. thermal, states are considered, the above does not hold. The study of systems at finite temperature possesses a higher level of difficulty. Most research has been performed in systems with additional symmetry, namely in conformal field theory \cite{Calabrese:2004eu,Cardy:2014jwa} or free scalar field theory in two dimensions \cite{Herzog:2012bw}. In more recent years, the latter have also been studied in the context of the holographic duality, via the Ryu-Takayanagi conjecture \cite{Ryu:2006bv}. 

Unlike the entanglement entropy, the mutual information possesses the symmetry property by definition and it is a good measure of correlations when the composite system lies in a mixed state. In this work, extending the work in \cite{Srednicki:1993im}, we study systems of coupled harmonic oscillators and free scalar field theory at finite temperature and show that the mutual information obeys indeed an area law, similar to the area law obeyed by the entanglement entropy at vanishing temperature. The coefficient of this area law does not vanish at infinite temperature, but tends to a given finite value that can be attributed to classical correlations.

\setcounter{equation}{0}
\section{Harmonic Oscillators at Finite Temperature}
\label{sec:qm}

Let us first consider the case of two coupled harmonic oscillators at finite temperature. Without loss of generality, we consider the oscillator masses equal to one. The Hamiltonian reads
\begin{equation}
H = \frac{1}{2}\left[ {{p}^2 + \left({p^C}\right)^2 + {k_0}\left( {{x}^2 + \left({x^C}\right)^2} \right) + {k_1}{{\left( {{x^C} - {x}} \right)}^2}} \right] .
\label{eq:hamiltonian_original_coordinates}
\end{equation}
One can introduce the canonical coordinates, ${x_ \pm } \equiv \left( {{{x^C} \pm {x}}} \right) / {{\sqrt 2 }}$. Then, the Hamiltonian assumes the form of two decoupled oscillators
\begin{equation}
H = \frac{1}{2}\left( {p_+ ^2 + p_- ^2 + {\omega _ + }^2x_+ ^2 + {\omega _ - }^2x_- ^2} \right) ,
\label{eq:hamiltonian_normal_coordinates}
\end{equation}
where $\omega_\pm$ are the eigenfrequencies of the two normal modes, namely, $\omega_+ = \sqrt{k_0}$ and $\omega_- = \sqrt{k_0 + 2 k_1}$.
In is easy to show that the thermal density matrix of a single harmonic oscillator with eigenfrequency $\omega$ is given by
\begin{equation}
\rho \left( {x,x'} \right) = \frac{{2\sinh \frac{\omega }{{2T}}}}{{\sqrt {2\sinh \frac{\omega }{T}} }}\sqrt {\frac{\omega }{\pi }} {e^{ - \frac{{\omega \left( {{x^2} + x{'^2}} \right) \coth\frac{\omega }{T}}}{2}}}{e^{\frac{{\omega xx'}}{{\sinh \frac{\omega }{T}}}}} = \sqrt {\frac{\omega }{\pi }\left( {a + b} \right)} {e^{ - \frac{{a\left( {{x^2} + x{'^2}} \right)}}{2}}}{e^{ - bxx'}} ,
\label{eq:ho_density_2}
\end{equation}
where the quantities $a$ and $b$ are equal to $a \equiv \omega \coth \frac{\omega }{T}$, $b \equiv  - \omega \csch \frac{\omega }{T}$. It follows that the thermal density matrix of the system of the two coupled oscillators reads
\begin{multline}
\rho \left( {{x_ + },{x_ + }',{x_ - },{x_ - }'} \right) = \rho \left( {{x_ + },{x_ + }'} \right) \otimes \rho \left( {{x_ - },{x_ - }'} \right)\\
= \frac{{\sqrt {\left( {{a_ + } + {b_ + }} \right)\left( {{a_ - } + {b_ - }} \right)} }}{\pi }{e^{ - \frac{{{a_ + }\left( {{x_ + }^2 + {x_ + }{'^2}} \right) + {a_ - }\left( {{x_ - }^2 + {x_ - }{'^2}} \right)}}{2}}}{e^{ - {b_ + }{x_ + }{x_ + }'}}{e^{ - {b_ - }{x_ - }{x_ - }'}},
\end{multline}
where $a_\pm \equiv \omega_\pm \coth \frac{\omega_\pm }{T}$, $b_\pm \equiv  - \omega_\pm \csch \frac{\omega_\pm }{T}$.

It is a simple exercise with Gaussian integrals to integrate out the degrees of freedom of the oscillator described by the coordinate $x^C$, in order to find the reduced density matrix that describes the other oscillator. It reads
\begin{equation}
\rho \left( {{x},{x}'} \right) = \int {d{x^C}\rho \left( {{x},{x}',{x^C},{x^C}} \right)} = \sqrt {\frac{{\gamma  - \beta }}{\pi }} {e^{ - \frac{{\left( {{x}^2 + {x}{'^2}} \right)\gamma }}{2}}}{e^{{x}{x}'\beta }} ,
\label{eq:two_osc_reduced}
\end{equation}
where
\begin{equation}
\gamma  - \beta  = 2\frac{{\left( {{a_ + } + {b_ + }} \right)\left( {{a_ - } + {b_ - }} \right)}}{{ {{a_ + } + {a_ - } + {b_ + } + {b_ - }} }} , \quad \gamma + \beta = \frac{1}{2} \left( {{a_ + } + {a_ - } - {b_ + } - {b_ - }} \right) .
\end{equation}

Similarly to the ground state analysis, it can be shown that the eigenfunctions of the above density matrix are
\begin{equation}
{f_n}\left( x \right) = {H_n}\left( {\sqrt \alpha  x} \right){e^{ - \frac{{\alpha {x^2}}}{2}}} ,
\label{eq:two_eigenfunctions}
\end{equation}
where
\begin{equation}
\alpha \equiv \sqrt {{\gamma ^2} - {\beta ^2}} .
\end{equation}
The corresponding eigenvalues are
\begin{equation}
{p_n} = \left( {1 - \frac{\beta }{{\gamma  + \alpha }}} \right){\left( {\frac{\beta }{{\gamma  + \alpha }}} \right)^n} \equiv \left( {1 - \xi } \right){\xi ^n} ,
\end{equation}
where
\begin{equation}
\xi \equiv \frac{\beta }{{\gamma  + \alpha }} = \frac{{1 - \sqrt {\frac{{\gamma  - \beta }}{{\gamma  + \beta }}} }}{{1 + \sqrt {\frac{{\gamma  - \beta }}{{\gamma  + \beta }}} }} .
\end{equation}

It is now straightforward to calculate the entanglement entropy. It equals
\begin{equation}
S_A =  - \ln \left( {1 - \xi } \right) - \frac{\xi }{{1 - \xi }}\ln \xi .
\label{eq:two_osc_SA}
\end{equation}
The symmetry of the problem ensures that $S_{A^C} = S_A$, which further implies that
\begin{equation}
I\left( {A:{A^C}} \right) = 2 {S_A} - S_{\textrm{th}} .
\end{equation}
The thermal entropy $S_{\textrm{th}}$ of the system is simply the sum of the thermal entropy of the two normal modes, namely
\begin{equation}
{S_{\textrm{th}}} =  - \ln \left( {1 - {e^{ - \frac{\omega_+ }{T}}}} \right) + \frac{\omega_+ }{T}\frac{1}{{{e^{\frac{\omega_+ }{T}}} - 1}} - \ln \left( {1 - {e^{ - \frac{\omega_- }{T}}}} \right) + \frac{\omega_- }{T}\frac{1}{{{e^{\frac{\omega_- }{T}}} - 1}} .
\label{eq:two_thermal_entropy}
\end{equation}

An interesting fact is that although $S_A$ and ${S_{\textrm{th}}}$ at large temperatures diverge as $\ln T$ and $2 \ln T$ respectively, the mutual information tends to a finite value. One can show that
\begin{equation}
\mathop {\lim }\limits_{T \to \infty } I\left( {A:{A^C}} \right) = \frac{1}{2}\ln \frac{{{{\left( {{k_0} + {k_1}} \right)}^2}}}{{{k_0}\left( {{k_0} + 2{k_1}} \right)}} .
\label{eq:2_I_highT_expansion}
\end{equation}
This is unlike the usual behaviour in qubit systems, where the mutual information vanishes at infinite temperature. The reason for this is the fact that the Hilbert space of each subsystem is infinite dimensional. In qubit systems, the Hilbert spaces are finite dimensional. Let $d_A$ be the dimension of the Hilbert space of subsystem $A$ and similarly $d_{A^C}$ the dimension of the Hilbert space of the complementary subsystem. Then, at the infinite temperature limit the reduced density matrices tend to $\rho_A = I_{d_A} / d_A$ and $\rho_{A^C} = I_{d_{A^C}} / d_{A^C}$, respectively. The overall density matrix tends to $\rho = I_{d_A d_{A^C}} / \left( d_A d_{A^C} \right)$. It follows that the mutual information tends to $I\left( {A:{A^C}} \right) = \ln d_A + \ln d_{A^C} - \ln \left( d_A d_{A^C} \right) = 0 $. However, in our case the Hilbert spaces are infinite dimensional. This mechanism enforces the cancellation of the terms, which are proportional to $\log T$, but there is a finite remnant left.

The mutual information captures both classical and quantum correlations between the considered subsystems. This remnant turns out to be exactly equal to what would have been obtained had one considered a thermal ensemble of classical pairs of coupled oscillators. Therefore, this remnant should be attributed to classical correlations solely. More specifically, assume a single oscillator at a state with energy $E$. Then, the probability of finding this oscillator at position $x$ is simply $p \left( x \right) \sim 1 / v \left( x ; E \right)$, where $v$ is the velocity, which can be calculated by conservation of energy. Considering a thermal ensemble, the probability of finding the oscillator at position $x$ is simply
\begin{equation}
p \left( x \right) \sim \int_0^\infty {dE e^{- \beta E} / v \left( x ; E \right)}.
\end{equation}
This probability distribution turns out to be Gaussian
\begin{equation}
{p_{{\rm{can}}}}\left( {x;\omega ,T} \right) = \int_{\frac{1}{2}{\omega ^2}{x^2}}^\infty  {p\left( E \right){p_E}\left( x \right)dE}  = \frac{\omega }{{\sqrt {2\pi T} }}{e^{ - \frac{{{\omega ^2}{x^2}}}{{2T}}}} ,
\end{equation}
It is simple to apply the above to the two normal modes of the system and find the probability distributions $p\left( {{x},{x^C};T} \right) = {p_{{\rm{can}}}}\left( {\frac{{{x} + {x^C}}}{{\sqrt 2 }};{\omega _ + },T} \right){p_{{\rm{can}}}}\left( {\frac{{{x} - {x^C}}}{{\sqrt 2 }};{\omega _ - },T} \right)$, as well as $p\left( {{x};T} \right) = \int {p\left( {{x},{x^C};T} \right)d{x^C}}$ and $p\left( {{x^C};T} \right) = \int {p\left( {{x},{x^C};T} \right)d{x}}$. It turns out that the mutual information corresponding to these distributions is finite, temperature independent and exactly equal to the infinite temperature limit of the quantum mutual information.

The above procedure can be easily generalized to a system of an arbitrary number of coupled harmonic oscillators. In the following we consider a system of $N$ oscillators, described by coordinates $x_i$ and as subsystem $A^C$ the first $n$ of those. Without loss of generality, we assume that all oscillators have unit mass. The Hamiltonian reads
\begin{equation}
H = \frac{1}{2}\sum\limits_{i = 1}^N {{p_i}^2}  + \frac{1}{2}\sum\limits_{i,j = 1}^N {{x_i}{K_{ij}}{x_j}} .
\end{equation}
The matrix $K$ is symmetric and it has positive eigenvalues, so that the above Hamiltonian describes an oscillatory system around a stable equilibrium position. Performing an orthogonal transformation to the normal coordinates, writing down the thermal density matrix for each mode and then performing the inverse orthogonal transformation back to the original coordinates, one may find that the thermal density matrix of this system is
\begin{equation}
\rho \left( {{\bf{x}},{\bf{x}}'} \right) = \sqrt {\frac{{\det \left( {a + b} \right)}}{{{\pi ^N}}}} {e^{ - \frac{{{x^T}ax + x{'^T}ax'}}{2}}}{e^{ - {x^T}bx'}} ,
\end{equation}
where the matrices $a$ and $b$ are defined as
\begin{equation}
a = \sqrt K \coth\frac{{\sqrt K }}{T},\quad b =  - \sqrt K \csch\frac{{\sqrt K }}{T}.
\label{eq:N_ab_defs}
\end{equation}

Now it is a matter of algebra to integrate out the first $n$ degrees of freedom in order to find the reduced density matrix describing subsystem $A$. We write any $N \times N$ matrix $M$ in block form as
\begin{equation}
M = \left( {\begin{array}{*{20}{c}}
{{M_A}}&{{M_B}}\\
{M_B^T}&{{M_C}}
\end{array}} \right) ,
\end{equation}
where $M_A$ is an $n\times n$ matrix, $M_C$ is an $\left( N - n \right) \times \left( N - n \right)$ matrix and finally $M_B$ is an $n\times \left( N - n \right)$ matrix. Then, it turns out that
\begin{equation}
{\rho_A}\left( {{x},{x}'} \right) = \sqrt {\frac{{\det \left( {\gamma  - \beta } \right)}}{{{\pi ^{N - n}}}}} {e^{ - \frac{{{x}^T\gamma {x} + {x}{'^T}\gamma {x}'}}{2}}}{e^{{x}^T\beta {x}'}} ,
\end{equation}
where
\begin{align}
\gamma  &= a_C - \frac{1}{2}\left( {{a_B^T} + b_B^T} \right){\left( {a_A + b_A} \right)^{ - 1}}\left( {a_B + b_B} \right) , \label{eq:N_gamma_def}\\
\beta  &=  - b_C + \frac{1}{2}\left( {{a_B^T} + b_B^T} \right){\left( {a_A + b_A} \right)^{ - 1}}\left( {a_B + b_B} \right) . \label{eq:N_beta_def}
\end{align}
Having obtained this expression, one may proceed to calculate the spectrum of this reduced density matrix, and, thus, the entanglement entropy and the mutual information, in exactly the same way as in the ground state case \cite{Katsinis:2017qzh}. The spectrum of ${\rho_A}$ reads
\begin{equation}
{p_{{n_{n + 1}}, \ldots ,{n_N}}} = \prod\limits_{i = n + 1}^N {\left( {1 - {\xi _i}} \right)\xi _i^{{n_i}}} ,\quad {n_i} \in \mathbb{Z} ,
\end{equation}
where the quantities $\xi_i$ are given by
\begin{equation}
{\xi _i} = \frac{{{\beta _{Di}}}}{{1 + \sqrt {1 - \beta _{Di}^2} }}
\label{eq:N_xi_of_beta}
\end{equation}
and $\beta _{Di}$ the eigenvalues of the matrix $\gamma^{-1}\beta$. It follows that the entanglement entropy equals
\begin{equation}
S = \sum\limits_{j = n + 1}^N {\left( { - \ln \left( {1 - {\xi _j}} \right) - \frac{{{\xi _j}}}{{1 - {\xi _j}}}\ln {\xi _j}} \right)} .
\label{eq:N_S_of_xi}
\end{equation}

\setcounter{equation}{0}
\section{Free Scalar QFT}
\label{sec:qft}
Following \cite{Srednicki:1993im}, one may use the formulae derived in section \ref{sec:qm} to study the mutual information in free scalar quantum field at finite temperature. The Hamiltonian reads
\begin{equation}
H = \frac{1}{2}\int {{d^3}x\left[ {{\pi ^2}\left( {\vec x} \right) + {{\left| {\vec \nabla \varphi \left( {\vec x} \right)} \right|}^2} + {\mu^2}\varphi {{\left( {\vec x} \right)}^2}} \right]} .
\label{eq:discretize_hamiltonian}
\end{equation}
For this purpose one needs to discretize the degrees of freedom, first expanding the field in real spherical harmonics
\begin{align}
{\varphi _{lm}}\left( x \right) &= x\int {d\Omega {Y_{lm}}\left( {\theta ,\varphi } \right)\varphi \left( {\vec x} \right)} ,\label{eq:discretize_def1}\\
{\pi _{lm}}\left( x \right) &= x\int {d\Omega {Y_{lm}}\left( {\theta ,\varphi } \right)\pi \left( {\vec x} \right)} ,\label{eq:discretize_def2}
\end{align}
and then introducing a lattice of spherical shells with radii $x = ja$, where $j \in \mathbb{N}$ and $1 \le j \le N$. The latter discretization introduces a UV cutoff to our system that equals $1/a$, while the overall size of the system sets an IR cutoff equal to $1/({Na})$. One can show that the coordinates $\varphi_{lm,j} = {\varphi _{lm}}\left( {ja} \right)$ and momenta $\pi _{lm,j} = a {\pi _{lm}}\left( {ja} \right)$ obey canonical commutation relations. The discretized Hamiltonian reads
\begin{equation}
H = \frac{1}{{2a}}\sum\limits_{l,m} {\sum\limits_{j = 1}^N {\left[ {{\pi _{lm,j}}^2 + {{\left( {j + \frac{1}{2}} \right)}^2}{{\left( {\frac{{{\varphi _{lm,j + 1}}}}{{j + 1}} - \frac{{{\varphi _{lm,j}}}}{j}} \right)}^2} + \left( {\frac{{l\left( {l + 1} \right)}}{{{j^2}}} + {\mu^2}{a^2}} \right){\varphi _{lm,j}}^2} \right]} } .
\end{equation}

Different $l$ and $m$ indices are not coupled. Furthermore, $m$ does not enter explicitly in the Hamiltonian. Therefore, the problem can be split to an infinite number of independent sectors, identified by index $l$, each one containing $2 l + 1$ identical independent subsectors. Thus, the entanglement entropy and the mutual information can be calculated by the series
\begin{equation}
S\left( {N,n} \right) = \sum\limits_l {\left( {2l + 1} \right){S_{l}}\left( {N,n} \right)} , \quad I\left( {N,n} \right) = \sum\limits_l {\left( {2l + 1} \right){I_{l}}\left( {N,n} \right)} ,
\label{eq:qft_entropy_series}
\end{equation}
where ${S_{l}}\left( {N,n} \right)$ and ${I_{l}}\left( {N,n} \right)$ is the entanglement entropy and the mutual information of a system described by the Hamiltonian
\begin{equation}
H_l = \frac{1}{{2a}} {\sum\limits_{j = 1}^N {\left[ {{\pi _{l,j}}^2 + {{\left( {j + \frac{1}{2}} \right)}^2}{{\left( {\frac{{{\varphi _{l,j + 1}}}}{{j + 1}} - \frac{{{\varphi _{l,j}}}}{j}} \right)}^2} + \left( {\frac{{l\left( {l + 1} \right)}}{{{j^2}}} + {\mu^2}{a^2}} \right){\varphi _{l,j}}^2} \right]} } .
\label{eq:qft_hamiltonian_l}
\end{equation}
The latter contains a finite number of degrees of freedom and thus, ${S_{l}}\left( {N,n} \right)$ and ${I_{l}}\left( {N,n} \right)$ at finite temperature $T$ can be calculated using the formulae derived in section \ref{sec:qm}.

For large $l$, the matrix describing the $N$ oscillators is dominated by its diagonal elements. As a result at this limit the system is almost disentangled. As a consequence, it can be shown that the series \eqref{eq:qft_entropy_series} is converging \cite{Srednicki:1993im}.

\begin{figure}[h]
\centering
\begin{picture}(100,37)
\put(1.5,2.5){\includegraphics[width = 0.45\textwidth]{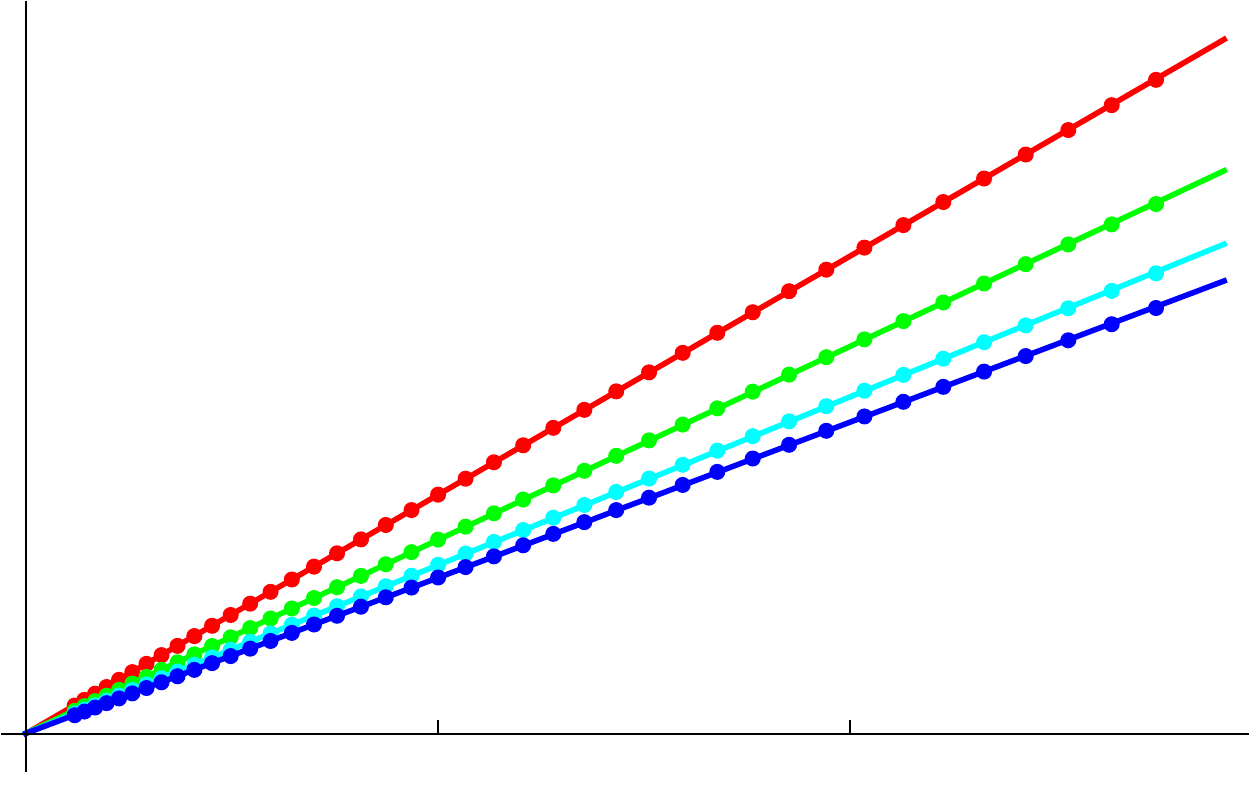}}
\put(51.5,2.5){\includegraphics[width = 0.45\textwidth]{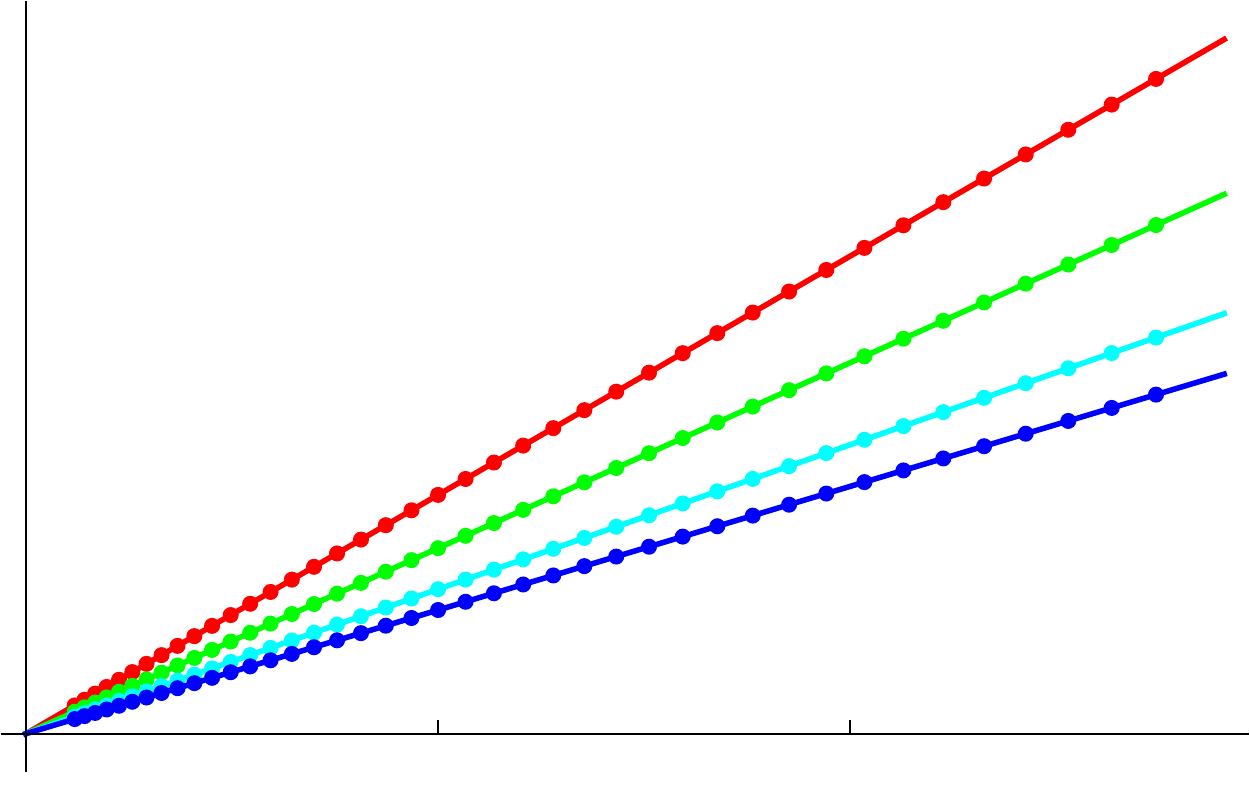}}
\put(56,18){\includegraphics[width = 0.06\textwidth]{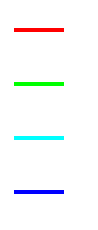}}
\put(61,19.5){$T\to \infty$}
\put(61,22.75){$T=1/a$}
\put(61,26){$T=1/(2a)$}
\put(61,29.25){$T=0$}
%
\put(21,34){$\mu = 0$}
\put(71,34){$\mu = 1 / a$}
\put(1.75,31.5){$I$}
\put(51.75,31.5){$I$}
\put(46.75,3.75){$\frac{R^2}{a^2}$}
\put(96.75,3.75){$\frac{R^2}{a^2}$}
\put(30.5,1.5){$\frac{N^2}{2}$}
\put(80.5,1.5){$\frac{N^2}{2}$}
\put(15.5,1.5){$\frac{N^2}{4}$}
\put(65.5,1.5){$\frac{N^2}{4}$}
\end{picture}
\caption{The mutual information as function of the size of the entangling sphere}
\label{fig:MI}
\end{figure}
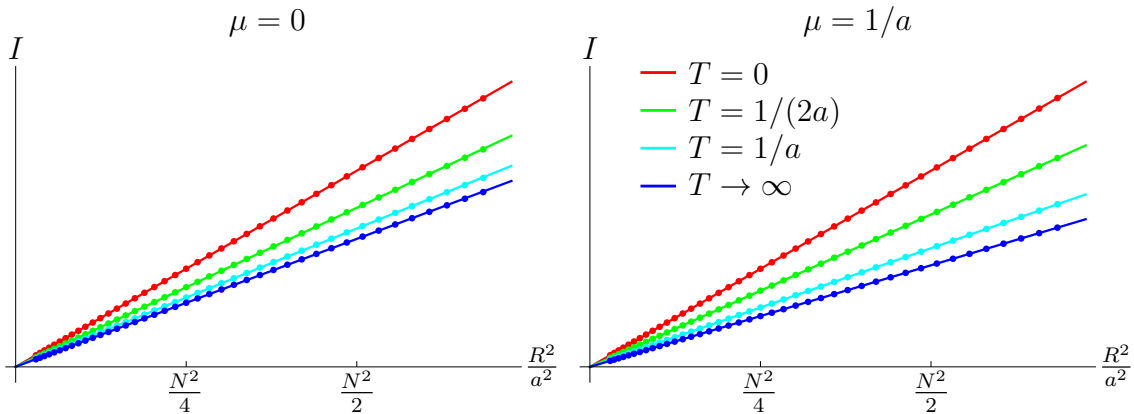

Figure \ref{fig:MI} shows the dependence of the mutual information on the size of the entangling sphere, both in the cases of a massless scalar field (left) and a massive one with $\mu a = 1$ (right). The numerical calculation of the eigenvalues of the relevant matrices has been performed with the help of Wolfram's Mathematica for $N=60$. It is evident that the mutual information is proportional to the area of the entangling sphere. In the case of the massless scalar field, at vanishing temperature we find that $I \simeq 0.59 R^2 / a^2$, which agrees with the result of \cite{Srednicki:1993im}. The coefficient of the area law term is a decreasing function of the temperature. However, it does not vanish as the temperature goes to infinity. It rather reaches an asymptotic finite value. In the case of the massless field, this value is approximately $I \simeq 0.38 R^2 / a^2$.

\setcounter{equation}{0}
\section{Discussion}
\label{sec:discussion}

The entanglement entropy in massless scalar field theory at its ground state was calculated for a spherical entangling surface in a classical work by Srednicki \cite{Srednicki:1993im}. It was found that the entanglement entropy is proportional to the area and not to the volume of the sphere. This property resembles the well-known property of the black hole entropy and motivates the investigation of new paths in the understanding of the nature of the gravitational interaction. More specifically, if the black hole entropy can be attributed to quantum entanglement entropy, either totally or partially, the gravitational interaction may be understood as a statistical force attributed to quantum entanglement statistics. Recent results from holographic theories also support such an interpretation. The ``area law'' behaviour of entanglement entropy also holds at massive scalar field theory, where apart from the numerical calculation similar to that in \cite{Srednicki:1993im}, analytical perturbative methods can be applied in order to calculate the entanglement entropy \cite{Katsinis:2017qzh}.

So, from a quantum mechanical point of view, is it possible to understand the underlying cause of this behaviour of entanglement entropy? In massive field theory, and more specifically at the limit of a very large mass, the area law behaviour can be considered as a consequence of locality. At this limit, the correlations are dumped exponentially fast, and, thus, only correlation between adjacent degrees of freedom are significant. It follows that the entanglement entropy should be proportional to the number of neighbouring pairs of degrees of freedom that get separated by the entangling surface. These are obviously proportional to the area of the entangling surface. But how does this behaviour insist even at the massless case? The underlying reason for this behaviour is the symmetry property of the entanglement entropy. Let $A$ be the considered subsystem and $A^C$ its composite. Then, when the composite system lies at any pure state, it holds that $S_A = S_{A^C}$. This excludes the existence of a volume term, as this should be proportional to the volume of subsystem $A$ \emph{and simultaneously} to the volume of its complement $A^C$, which is obviously impossible. The symmetry property enforces the entanglement entropy to depend solely on the geometric features of the two subsystems that they have in common, i.e. the geometry of the entangling surface.

The symmetry property does not hold whenever the composite system lies in a mixed state. Actually, this is the very reason the entanglement entropy is not a good measure of entanglement for such configurations. In such cases, the simplest extension of entanglement entropy, which is a good measure of the correlations between the two subsystems is the mutual information. This quantity possesses the symmetry property by definition. It follows that an ``area law'' behaviour for the mutual information in field theory, even for mixed states, e.g. thermal states, should not be considered surprising. It turns out that the complexity of the calculation of the mutual information in field theory at finite temperature is similar to that of the entanglement entropy at the ground state. We show that indeed the mutual information obeys an area law.

The coefficient which connects the area to the mutual information has an interesting dependence on the temperature. It is in general a decreasing function of temperature. At the limit of infinite temperature, this coefficient does not vanish, but it rather asymptotically tends to a finite value. This coincides with the result  of the analogous calculation in a classical system of coupled oscillators.

\subsection*{Acknowledgements}
The research of and G.P. has received funding from the Hellenic Foundation for Research and Innovation (HFRI) and the General Secretariat for Research and Technology (GSRT), in the framework of the ``First Post-doctoral researchers support'', under grant agreement No 2595. The research of D.K. is co-financed by Greece and the European Union (European Social Fund - ESF) through the Operational Programme ``Human Resources Development, Education and Lifelong Learning'' in the context of the project ``Strengthening Human Resources Research Potential via Doctorate Research'' (MIS-5000432), implemented by the State Scholarships Foundation (IKY). The authors would like to thank M. Axenides and E. Floratos for useful discussions.

\end{document}